\documentclass[preprint2]{emulateapj}
\usepackage{lscape,graphicx}
\usepackage{rotating}
\usepackage{natbib}
\bibpunct{(}{)}{;}{a}{}{,} 

\shorttitle{Quark-nova ejecta}
\shortauthors{Ouyed and Leahy}

\begin{document}

\title{Dynamical and thermal 
 evolution of the quark-nova ejecta}

\author{Rachid Ouyed and Denis Leahy}

\affil{Department of Physics and Astronomy, University of Calgary, 
2500 University Drive NW, Calgary, Alberta, T2N 1N4 Canada}

\email{ouyed@phas.ucalgary.ca}

\begin{abstract}
We explore  the dynamical and thermal evolution of the
 ejected neutron star crust in a Quark-Nova explosion. Typical
  explosion energies and ejected crust masses result in relativistic ejection
  with Lorentz factors of a few to a few hundred.  The ejecta undergoes a rapid
  cooling and stretching resulting in  break up into many small pieces (clumps) when the
   ejecta is only $\sim 100$ km from the explosion site.  
  The number and size of the clumps depends
      on whether the breakup occurs in the liquid or solid phase.
       For these two cases, the clump number is $\sim 10^3$ (liquid phase) or $\sim 10^7$ (solid phase)
        and, at break up, are spherical  (size $\sim 10^4$ cm; liquid phase) or needle shaped ($\sim 10^4\times 10^2$ cm; solid phase).  
\end{abstract}

\section{INTRODUCTION}

In the quark-nova (QN) picture, (Ouyed et al. 2002;  Ker\"anen et al. 2005; hereafter
ODD and KOJ respectively) the 
core of a neutron star, that undergoes the phase transition to the quark phase, 
shrinks in a spherically symmetric fashion to a stable, more compact strange 
matter configuration faster than the overlaying material (the neutron-rich
hadronic envelope) can respond, leading to an effective core collapse.
The core of the neutron star is a few kilometers in radius initially, 
but shrinks to 1-2 km in a collapse time of about 0.1 ms
\cite{LBV}. The gravitational potential energy
released (plus latent heat of
phase transition) during this event is converted partly into internal energy  and partly into outward propagating shock waves
which impart kinetic energy to the material that eventually forms the
ejecta.

 There are three previously proposed mechanisms for  ejection of the outer layers
 of the neutron star (i.e. crust): 
  (i) Unstable baryon to quark combustion 
   leading to a shock-driven  ejection (Horvath \& Benvenuto 1988).
   More recent work, assuming realistic quark matter equations of state, argues for strong
deflagration (Drago et al. 2007)  that can expel surface material.  In these
models up to $10^{-2} M_{\odot}$ can be ejected.
These calculations focus on the microphysics and 
    not on the effect of the global state of the resulting quark core
     which collapses prior to complete combustion (KOJ), leading
     to conversion only of the inner core ($\sim$ 1-2 km) of the neutron star;
   (ii) Neutrino-driven explosion where the energy is deposited
  in a thin (the densest) layer at the bottom of the crust above a gap 
  separating it from the  collapsing core (KOJ).    
     For the neutrino-driven mechanism,
    the core bounce was neglected, and neutrinos emitted from the conversion to
strange matter  transported the energy into the outer
regions of the star, leading to heating and subsequent mass ejection. 
Consequently, mass ejection is limited to about $10^{-5}M_{\odot}$ (corresponding to the
crust mass below neutron drip density)  for compact quark cores of size (1-2) km;
 (iii) Thermal fireball driven ejection which we consider for the
 present study. The fireball is inherent to the properties of the quark star at birth.
   The birth temperature  was found
   to be of the order of 10-20 MeV since the collapse is adiabatic
    rather than isothermal (ODD; KOJ).  In this temperature regime the
    quark matter is in the superconducting Color-Flavor Locked (CFL)
     phase (Rajagopal\&Wilczek 2001)  where the photon emissivity
     dwarfs the neutrino emissivity (Vogt, Rapp, \& Ouyed 2004; Ouyed, Rapp, \& Vogt 2005).  
  The average photon energy is $\sim$3$T$ for a CFL temperature $T$:
  since the plasma frequency of CFL matter is $\hbar\omega\sim 23$ MeV (e.g. Usov 2001),
    the photon emissivity is highly attenuated as soon
as the surface temperature of the star cools
below $T_{\rm a} = \hbar\omega_{\rm p}/3\simeq 7.7$~MeV. To summarize,
 the thermal fireball is generated as the star cools from its birth temperature
 down to $\sim 7.7$ MeV.

 Here we focus on the case where a QN goes off
 in isolation, neglecting any interaction with the surroundings, 
 and the crust is ejected by a thermal fireball (i.e. case (iii) above). 
 The remainder of this  paper is presented as follows: In Sect. 2
  we discuss the energetics of the QN and the resulting fireball and ensuing
  crust ejection.
 Sect. 3   deals with the hydrodynamical
and thermal evolution of the ejecta. 
 We   conclude in Sect. 4.
 
 \section{Fireball generation and crust ejection}

The QN ejecta, which is the left-over crust of the parent neutron star,
is initially in the shape of a shell and is imparted with energy from the QN explosion.
This energy released during the QN explosion ($E_{\rm QN,53}$; in units of $10^{53}$ erg) 
is a combination of baryon to quark conversion energy and gravitational energy
release due to contraction.  Also, the rapid contraction creates a gap between the
surface of the collapsed quark matter core and the inner edge of the remaining hadronic
matter. 
Before entering the CFL phase at $T\sim 10$-$20$ MeV the QS
cools mainly by neutrino emission;  as shown in Ker\"anen et al. (2005)
 a significant fraction of neutrinos
escape through the crust. Once the QS core enters the CFL phase,
 the energy is released in photons  (see Vogt et al. 2004;
 Ouyed et al. 2005).   Then at that point   thermalization leads to the
creation of a $(e^+e^-)$ fireball with a temperature of 10-20 MeV.
 Thus the initial state for crust ejection is a QS core
 surrounded by a fireball in the gap ($\sim 1$ km) in turn surrounded by the initially
stationary neutron star crust. 

The general picture is that the fireball from the QN acts
as a piston at the base of the crust; i.e. the QS fireball
expands approximately adiabatically while pushing the overlaying crust, and cooling fairly rapidly.
Using the relativistic force equation, a simple estimate for the
timescale to accelerate the crust to Lorentz factor of a few is $\sim$ 0.1 to 1 millisecond.
 The force is due to the fireball pressure ($F= PA$ where $P=(a/3)T^4$,
  $T\sim 10$-20 MeV and $A$ is
the shell's area). The acceleration timescale is only a factor of a few smaller
 than the free-fall timescale. As the crust moves outward, the fireball
cools approximately adiabatically,
 with the temperature $T$ proportional to $V^{-1/3}$ with $V$ the volume between the
quark star surface and the inner edge of the crust. Initially the temperature
decreases slowly with $R(t)$ then later it decreases as $1/R(t)$ where $R(t)$ is the radial
 position of the crust. Furthermore,  the final kinetic
energy of the crust for a Lorentz factor of $10$ is about  $2\times 10^{50}$ erg
for a $10^{-4}M_{\odot}$ crust. Thus the energy needed to eject
the crust is less  than 1\% of fireball energy.

 The energy input to the crust
is mainly kinetic and only a small fraction is thermal since there is no strong shock in this scenario.
During the acceleration, the crust
can respond on timescale $\Delta R/c_s$ where $\Delta R$ is the crust thickness
 (about 1 km) and $c_s$ is the speed of sound in the crust. The majority
of the crust is degenerate with $c_s=c/\sqrt{3}$. The response timescale is
  $\sim 5\times 10^{-6}$ seconds, 
about $(1/100)$th of the acceleration timescale. Thus no strong shock
will be generated except in the outermost non-degenerate (i.e. low sound speed) layers
 (the mass contained in the non-degenrate layers is small $< 10^{-6}M_{\odot}$).
 The bulk of the crust is thus accelerated smoothly by the 
 pressure from the fireball.  In comparison,  a normal SN has the energy of core collapse
released into mostly non-degenerate matter.  Thus a strong shock
 is generated. In the QN case, most of the crust is highly degenerate with
 high sound speed, thus it can be accelerated without a strong shock.

 One can estimate the Lorentz factor\footnote{For simplicity we assume that the entire
ejected crust can be represented using a single Lorentz factor, density and temperature. 
 However,   a more realistic situation might involve 
  ejecta with a distribution of these variables. This
   can only be determined with calculation of
    the ejection process which requires complex simulations.} of the QN ejecta by,  
\begin{equation}
 \label{eq:etagamma}
 \Gamma_{\rm i}  = \frac{\eta E_{\rm QN}}{m_{\rm ejecta}c^{2}}
 \sim  50 \frac{\eta_{0.1}E_{\rm QN,53}}{m_{\rm ejecta, -4}}\ ,
\end{equation}
where  $\eta_{0.1}$ is the efficiency of energy transfer from the QN to
the ejecta's kinetic energy in units of 0.1, and the 
ejecta mass, $m_{\rm ejecta,-4}$ is given in units of $10^{-4}M_{\odot}$. 
Mass ejection in   the QN scenario depends on details of energy transfer to the crust.
 Best estimates of ejected mass from existing calculations are $10^{-5}M_{\odot}$-$10^{-2}M_{\odot}$.
We adopt a rounded, fiducial, ejected mass of $10^{-4}M_{\odot}$ for the remainder of this paper.
 Heavier mass ejection would lead to mildly or non-relativistic
  ejecta.

\section{Evolution of the Quark-Nova Ejecta}

For ejecta masses higher than $10^{-4}M_{\odot}$ the 
 corresponding initial density and size  (i.e. of the neutron star crust)  just before ejection are
  $\rho_0> 10^{10}$ g cm$^{-3}$ and $\Delta r_0 > 0.025 r_0$
  respectively (e.g. Datta et al. 1995), where $r_0$ is the radius of the
  parent neutron star before the QN explosion.  Also, the ejecta's birth
  temperature, $T_0$, is estimated to be of the order of 10 MeV.
Using general volume expansion ($V=V_{0}(r/r_{0})^{\alpha_{\rm V}}$), mass conservation implies that the density
of the ejecta will be $\rho = \rho_{0} (r/r_0)^{-\alpha_{\rm V}}$.
We define $\alpha_{\rm V}$ by a power law dependence
of $V$ on $r$  with different $\alpha_{\rm V}$ describing different physical
situations, e.g. constant volume, adiabatic, etc..
At this density and temperature the ejecta is relativistic degenerate, 
so that the Fermi energy evolves with radius as 
$\epsilon_{\rm F}\propto \rho^{1/3} \propto (\frac{r}{r_0})^{-\alpha_{\rm V}/3}$ (see Appendix \ref{appendixA} for details). 

The total internal energy, $U$, in the ejecta at birth can be estimated to be 
$U_0 \sim 5\times 10^{47}\  {\rm erg} \  m_{\rm ejecta, -4} \rho_{0,10}^{1/3}$. 
Here, the ejecta mass is given in units of $10^{-4}M_{\odot}$, its density
in units of $10^{10}$ g cm$^{-3}$, and $\mu_{\rm e}=2$ is the mean molecular weight
 dominated by the degenerate electrons.
The corresponding heat equation describing the thermal evolution of the ejecta can then be written as
\begin{equation}
\frac{d U}{dt} = -L_{\rm \gamma} - P\frac{dV}{dt}=-L_{\rm \gamma} - \frac{2}{3}U\frac{d\ln V}{dt}\ ,
\end{equation}
where all quantities are expressed in the ejecta's frame.
Since the ejecta is expanding relativistically, in its frame we can approximate
$dr = c dt$ with $r$ in this case being the distance
to the star from the ejecta. This allows the equation above to be recast into, 
\begin{equation}\label{eq:heat}
 r^{-2\alpha_{\rm V}/3}\frac{d (Ur^{2\alpha_{\rm V}/3})}{dr} = -\frac{L_{\rm \gamma}}{c}\ ,
\end{equation}

 \subsection{Relativistic expansion of a degenerate ejecta}
\label{appendixB}

We solve for the heat equation (Eq. \ref{eq:heat}) for the cases of 
 adiabatic and  isothermal expansion of the QN ejecta in the shape of a spherical
shell. The expansion is relativistic and assumed to occur at  at constant speed.
  The analysis  is  in the ejecta's frame. In the adiabatic case, 
  $L_{\gamma}=0$ which leads to 
 $U/U_0 = (r/r_0)^{-2\alpha_{\rm V}/3}$. Using the expression for $\epsilon_{F}$ (with $\epsilon_{F,0}=\epsilon_{F}(\rho_{0})$)
 given in Appendix A we arrive at 
 \begin{equation}
 \frac{1+\frac{5\pi^2}{12}(\frac{kT}{\epsilon_{\rm F}})^2}{1+\frac{5\pi^2}{12}(\frac{kT_0}{\epsilon_{\rm F,0}})^2}=(\frac{r}{r_0})^{-\alpha_{\rm V}/3} \ .
 \end{equation}
which shows that the temperature goes to zero in finite time corresponding to a radius
\begin{equation}
\frac{r}{r_0} = \left( 1 +  \frac{5\pi^2}{12}(\frac{kT_0}{\epsilon_{\rm F,0}})^2\right)^{3/\alpha_{\rm V}}\ .
\end{equation}
Clearly the radius is too close to the origin which means that in the 
adiabatic case the ejecta would cool almost immediately and would continue to expand
 at zero temperature, with $U=(3/5)N\epsilon_{\rm F}$.
  In reality the ejecta will solidify when it cools below the melting temperature of iron (see
  \S 2.2 below).

 In the isothermal case,  $T=T_0$,   the heat equation leads to 
 \begin{equation}
 \frac{U}{U_0}=(\frac{r}{r_0})^{-2\alpha_{\rm V}/3}  - A_{\alpha_{\rm V}} \left( (\frac{r}{r_0}) - (\frac{r}{r_0})^{-2\alpha_{\rm V}/3}\right) \ ,
 \end{equation}
with $A_{\alpha_{\rm V}} \sim \frac{6.11}{\frac{2\alpha_{\rm V}}{3} +1} (r_{0,6}^3 T_{0,10}^4/m_{-4}\rho_{10}^{1/3})$; $r_{0,6}$ is $r_0$ in units of $10^6$ cm and $T_{0,10}$
is $T_0$ in units of 10 MeV.
 The  internal energy $U$ goes to zero at a finite radius
 \begin{equation}
 \frac{r}{r_0} = \left( \frac{1}{A_{\alpha_{\rm V}}} +1 \right)^{1/(1+\frac{2\alpha_{\rm V}}{3})}\ ,
 \end{equation}
   which corresponds to $r\sim 1.15 r_0$ for $\alpha_{\rm v}\sim 9/4$.
 The corresponding  ejecta density at this radius is $\sim 10^{9}$ gm cm$^{-3}$.
  For $T_0  \sim 0.1$ MeV we find $r\sim 17 r_0$ for $\alpha_{\rm v}\sim 9/4$. with a 
  corresponding  ejecta density  $\sim  10^{8}$ gm cm$^{-3}$.

In summary, applying the heat equation to the relativistically 
expanding ejecta shows  an almost instantaneous loss of the internal energy. 
This mainly due to the $PdV$ work done by the ejecta as it expands or due to its rapidly
increase in area which leads to efficient cooling.  
  
 \subsection{Liquid to solid transition} \label{sec:solid_liquid}

 In the early stages the QN ejecta, with a temperature in the tens of MeV,
resembles a hot molten plasma.  Because it is degenerate
 it cools extremely rapidly (see Appendix A). 
  As the ejecta moves outwards it expands and cools undergoing a liquid
   to solid transformation  -- we will refer to the corresponding radius as the 
solidification radius $r_{\rm s}$.  
The state of ions (in this case iron nuclei) can be conveniently specified by the
Coulomb plasma parameter (e.g. Potekhin et al. 1999),
\begin{equation}
\Xi = \frac{9.26 Z^2}{T_{\rm keV}}\left(\frac{\rho_{8}}{A}\right)^{1/3}\simeq 1640 \frac{\rho_{8}^{1/3}}{T_{\rm keV}}\ ,
\end{equation}
 where $Z$ and $A$ are  the  ion charge number and  the atomic weight of iron.
  Since here $\Xi > 1$ prior to solidification, the shell material constitutes a strongly 
   coupled liquid.
  Solidification occurs for $\Xi > \Xi_{\rm m}=172$ which
  implies a solidification temperature of $T_{\rm s}\sim 9.5 \rho_{8}^{1/3}$ keV (see also de Blasio 1995).
 For example, using equation above and for $\alpha_{\rm v}\sim 9/4$, an ejecta born at $\sim 10$ MeV
  and $\rho_0 = 10^{10}$-$10^{11}$ g cm$^{-3}$, 
  will  cool to below $T_{\rm s} \sim 10$ keV and solidify (see \S 2.1) when it reaches a radius of  $\sim 3$-$10 r_{0}$.   The solidification  radius is then  $r_{\rm s}\sim 30$-$100$ km, 
at which point the ejecta's density is $\rho_{\rm s} > 10^{8}$ g cm$^{-3}$ (see Appendices \ref{appendixB}).  The corresponding  ejecta density at this radius is $\sim 10^{8}$ gm cm$^{-3}$.

 Previous studies of decompressed neutron star crust 
 show that the material heats up as it expands because the matter fuses into heavier elements and releases energy, which could prevent crystallization.  
The relevant calculations for expanding quark-nova ejecta has been  performed in detail in 
Jaikumar et al. (2007). That work considers
 r-process in quark-nova ejecta with and without $\beta$-decay heating. 
 In particular section 3 in that paper discusses the decompression of the ejecta. 
 As can be seen from Figures 5 and 6 in Jaikumar et al. (2007) the ejecta does not reheat above 10 keV (see peak in temperature at time of a few milliseconds). 
 The temperature prior to $\beta$-decay reheating is  low
enough for crystallization and break-up to occur.
However, $\beta$-decay reheating which peaks at a few milliseconds after ejection 
 (see figures 5 and 6 in Jaikumar et al. 2007) might re-melt the crystallized ejecta. Even so, the melting temperature and  the temperature of the re-heated ejecta may be similar: this would lead to 
crystallization at some later time.  
Finally, we note that those calculations  did not include
radiative losses so they provide an upper limit on the ejecta 
reheating temperature, thus increasing the likelihood of crystallization.

 \subsection{Ejecta Breakup} \label{sec:breakup}

Regardless of when crystallization occurs,  the relativistic expansion causes rapid breakup into small chunks 
  because of the inability of causal communication laterally in the shell.
   Specifically, as the spherical shell expands radially outwards relativistically,
 it also stretches laterally relativistically\footnote{The lateral separation velocity
  for 2 points separated by $\Delta \theta$ is $c\Delta \theta$ (see Appendix \ref{appendixD}).}.  Since the lateral expansion of the matter
  in the shell is  limited by the speed of sound, $c_{\rm s}$, 
   adjacent patches of the shell separate  leading to breakup 
   into small clumps. 
   
   The breakup depends on the existence of surface tension in the expanding shell.
For liquid or solid iron (our case), inter-ionic forces (mediated by the electrons) provide the tension.
 Such a tension does not exist in a gas. E.g., for SN ejecta where the 
 shock  accelerates the gas beyond the sound speed  the lack
 of surface tension means the particles simply expand away from each other. 
 An analogy is that of a bursting balloon versus expanding dust.
 For the innermost SN ejecta the density may be high enough that it behaves
 like a liquid rather than a gas. In that case, it is likely the sound
speed (which increases inward) is higher than the expansion speed in the inner parts (which
decreases inward;  $v\propto r$). If the SN ejecta is still a liquid at the radius 
  where the expansion speed exceeds the speed of sound then it will be subject
 to this lateral breakup mechanism.

   The size of the clumps depends
      on whether the breakup occurs in the liquid or solid phase.
       We discuss the solid phase case first; the liquid
        case differs in having a much larger breakup strain $\psi$.
    As the ejecta solidifies at $r_{\rm s}$, a strain rapidly builds up inside
 as the ejecta continues to expand relativistically.
Defining the breaking strain
of solid iron to be $\psi\sim 10^{-3}$ (e.g. Halliday \& Resnick \S 13.6) then the ejecta
break-up occurs at roughly a radius of $r_{\rm b} \approx (1+\psi) r_{\rm s}$ (see Appendix \ref{appendixD}).
The typical chunk size at birth is $c_{\rm s}\Delta t$
where $\Delta t = (r_{\rm b}-r_{\rm s})/c$, or, 
\begin{equation}
 \Delta r_{\rm c} \sim  186 \ {\rm cm}\  \zeta_{\rm  b} r_{\rm b,100}\ ,
\end{equation}
where we defined $\zeta_{\rm b}=\psi_{-3}\rho_{\rm b,8}^{1/6}$.
Here $\psi$ and the breakup radius are given in units
of $10^{-3}$ and 100 km respectively, while the density
at solidification radius is in units of $10^{8}$ g cm$^{-3}$.

\begin{table}
\caption{Clump/chunk properties at breakup for Liquid (L) and Solid (S; see Appendix \ref{appendixD})  cases}
\begin{tabular}{|c|ccccc|}\hline
   & $\psi$ &  $\Delta r_{\rm c}$ (cm) & $\theta_{\rm c}$ (rad) & $m_{\rm c}$ (gm)& $N_{\rm c}$ \\\hline
  S & $10^{-3}$ &  190   &  $2\times 10^{-5}$    & $1.7\times 10^{19}$   & $2.5\times 10^{7}$          \\\hline
     L & $10^{-1}$ &  $1.9\times 10^{4}$   &  $2\times 10^{-3}$    & $1.7\times 10^{23}$   & $2.5\times 10^{3}$          \\\hline  
\end{tabular}
\end{table}

 Table 1 lists the properties of the clumps/chunks 
  for the two cases of breakup occurring in early liquid or later solid phase.
   If the shell breaks into clumps during the liquid phase, there are two
    possibilities: 
    \begin{itemize}
    
    \item (i)  the chunks will breakup into smaller pieces when they solidify.
    Because of the degeneracy of the ejecta, the cooling is so rapid
    that even if initial breakup occurs in the liquid phase it will
    be immediately followed by another breakup governed by the smaller 
     breaking strain $\psi$ of the solid.  Since the breakup radius
     is hardly changed the solid chunk size should be independent
      of whether clumping first took place during the liquid phase or not.
    
    \item
     (ii) the ambient temperature remains high enough to keep the clumps
      liquid as they expand.  We note that as the density drops
      due to expansion, the crystallization temperature also drops
       ($T_{\rm m} \propto r^{-2/3}$ since $\rho\propto r^{-2}$). There is
        a delicate competition between the dropping ambient temperature and the
        dropping crystallization temperature. 
        
        \end{itemize}
        
        In summary, if during the radial expansion, 
         the ambient temperature drops below $T_{\rm s}$, then the liquid will crystallize.
          We adopt case (i) as our fiducial case. However most of our results are
          the same for both cases and we note whenever case (ii) gives a different outcome.
           In what follows we use the term chunks to refer to either the
             small solid iron pieces or the larger liquid clumps.

\subsection{Expansion of broken ejecta}     
     
Beyond the break-up radius, $r_{\rm b}$, the chunks  remain in contact with each other
 within the relativistically expanding ejecta as a whole.
This is because the pieces expand in volume, filling up the space between them,
and causing the density of each piece to continuously decrease,
until they reach the zero pressure iron
density ($\rho_{\rm Fe}\sim 10$ g cm$^{-3}$), at which point they stop expanding.
If we define a filling factor to be, $f_{\rm b}=1$ at $r_{\rm b}$, 
then it will remain unity until some separation radius $r_{\rm sep.}$.
This radius of separation can be found from 
equation (\ref{eq:rhoshell}).  With $\rho=\rho_{\rm Fe}$ and our fiducial values this radius is
\begin{equation}
 \frac{r_{\rm sep.}}{r_{0}}\sim 1300 \ , 
\end{equation}
which implies that the chunks remain closely packed until the ejecta as a whole 
reaches $r_{\rm sep.}\sim 10^9$ cm.  This is an upper limit for the separation
radius since we neglected the effects of the ions  (see Appendix B).

An estimation of the cross-sectional area extended by each chunk
 just before separation is given by $N_{\rm c} \Delta r_{\rm c, sep.}^{2} = \theta_{\rm B}^{2} r_{\rm sep.}^{2}$, 
or, $\Delta r_{\rm c, sep.}\sim 3\times 10^{4}\ {\rm cm}\ \zeta_{\rm b}$ (see Appendix \ref{appendixD} for details).
Furthermore, their length can be found, from equation (\ref{eq:drshell}),
to be $\Delta r_{\rm ejecta, sep.}\sim 5\times 10^{6}\ {\rm cm}\ m_{\rm ejecta,-4}^{1/4}$.
Hence, we expect the chunk's length is roughly a hundred times its width
 for case (i) or equal to its width for case (ii).
In the observer's frame the ratio of length to width will be 
contracted by a factor of  $1/\Gamma_{\rm i}$.  
 For case (i), the chunks thus resemble  
what we will refer to as ``iron needles", reminiscent of the subjet model of
Toma et al. (2005) in the context of GRBs.  Finally, as the chunks expand radially outwards
 to a radius $r_{\rm out} > r_{\rm sep}$, we can associate a filling factor
$f_{\rm out} =  \frac{r_{\rm sep}^{2}}{r_{\rm out}^{2}}$.

\section{Summary and conclusion}
\label{sec:conclusion}

In this paper we investigate the thermal and dynamic evolution
 of a relativistically expanding iron-rich shell from a QN explosion.
  The QN produces a photon fireball which acts as piston
   to eject and accelerate the crust of the parent neutron star to
   mildly relativistic speeds.  We find that the shell rapidly cools
    and breaks up into numerous ($10^3$ to $10^7$) chunks because of the rapid lateral expansion.
     Breakup may occur while the material is in liquid or solid
      phase: if liquid the chunks are nearly spherical; if solid they are 
       needle shaped moving parallel to their long axis. 

Although the presented model is based on physical arguments, 
most of these are in reality more complicated and so would require more detailed studies
 and the help of numerical simulations.
For example,  the process of clumping, crystallization, and breakup
       of the ejecta,  would require better knowledge of the
       ambient conditions surrounding the ejecta. The astrophysical implications
        will be discussed elsewhere.

\begin{acknowledgements}
 This work is supported by an operating grant from the Natural Research Council of Canada (NSERC).
\end{acknowledgements}

\appendix

\section{Fermi gas}
\label{appendixA}

 The Fermi energy  for a relativistic and a non-relativistic gas
  is $\epsilon_{\rm F} \simeq 1.413\times 10^{-5}\ {\rm erg}\ \rho_{10}^{1/3}\simeq 8.8\ {\rm MeV}\ \rho_{10}^{1/3}$ and $\epsilon_{\rm F} \simeq 1.219\times 10^{-4}\ {\rm erg}\ \rho_{10}^{2/3}\simeq 76.2\ {\rm MeV}\ \rho_{10}^{2/3}$ respectively  (e.g. Shapiro\&Teukolsky 1983; p24).
   The internal energy, $u$, per particle is
\begin{equation}\label{eqn:apndx_u}
u = \frac{3}{5}\epsilon_{\rm F}\left[1 + \frac{5\pi^2}{12}\left(\frac{kT}{\varepsilon_{\rm F}}\right)^2\right] \,
\end{equation}
with a corresponding pressure $P=(2/3)(u/v)$ where $v$ is the volume per particle.
 In the  relativistic degenerate regime, $P = \kappa_{\rm r}\rho^{4/3}$ with 
 $\kappa_{\rm r} = 1.244\times 10^{15} \mu_{\rm e}^{-4/3}$;
  in the non-relativistic regime  $P = \kappa_{\rm nr}\rho^{5/3}$ with 
 $\kappa_{\rm nr} = 9.91\times 10^{12} \mu_{\rm e}^{-5/3}$. The
 corresponding sound speed is $c_{\rm s,r}\simeq 2.57\times 10^{7}\rho^{1/6}$ cm s$^{-1}$ and
  $c_{\rm s,nr}\simeq 2.26\times 10^{6}\rho^{1/3}$ cm s$^{-1}$, respectively.
  The transition from relativistic to non-relativistic degeneracy occurs  at density 
   $\rho_{\rm tr}\simeq 2\times 10^{6}$ g cm$^{-3}$.
  Finally, the transition from Fermi-Dirac to Boltzmann statistics occurs at the degeneracy temperature
$kT\sim \epsilon_{\rm F}$.

\section{Evolution of ejecta thickness and density}
\label{appendixC}

We approximate the pressure in the ejecta by degenerate electron gas pressure neglecting
 the contribution from the non-relativistic ions. 
 The ejecta thickness increases in time at the speed of sound $d(\Delta r) = c_{\rm s} dt$
 with $dt=dr/c$ in the ejecta's frame.
 It is straightforward to show that the ejecta's density quickly drops below
  the transition density  $\rho_{\rm tr}\sim 2\times 10^{6}$ g cm$^{-3}$ so that
  most of the ejecta thickness expansion occurs during the non-relativistic degenerate phase.
   That is,  $d(\Delta r)\sim 2.26\times 10^{6}\rho^{1/3} dt$
  with $\rho = m_{\rm ejecta}/(4\pi r^{2} \Delta r)$. Combining these equation gives
   \begin{equation}
   \frac{\Delta r}{\Delta r_{0}}\simeq \left( 1 + 355 B \left((\frac{r}{r_0})^{1/3}-1\right) \right)^{3/4}\ ,
   \end{equation}
   where $B = m_{\rm ejecta,-4}^{1/3}r_{0,6}^{1/3}/\Delta r_{0,4}^{4/3}$ with the ejecta's
   thickness at birth $\Delta r_{0,4}$ given in units of $10^{4}$ cm, 
    corresponding to $\rho_{0}\sim 10^{10}$ g cm$^{-3}$ (e.g. Datta et al. 1995).
   For $r>>r_0$,
   \begin{equation}\label{eq:drshell}
   \Delta r \sim 2.6\times 10^{4}\ {\rm cm}\ m_{\rm ejecta,-4}^{1/4} r^{1/4}\ .
   \end{equation}
   Mass conservation then gives (for $r>>r_0$),
   \begin{equation}\label{eq:rhoshell}
   \frac{\rho}{\rho_{0}}\sim \frac{10^{-2}}{B^{3/4}}
\left( \frac{r_0}{r} \right)^{9/4}\ .
   \end{equation}

\section{Ejecta Break-Up}
\label{appendixD}

A spherical shell expanding radially at a speed of light gives 
 2 points separated by $\Delta \theta$ a lateral separation
 velocity $c\Delta \theta$.  This causes a strain to rapidly build up
  leading to breakup. Here we discuss the case that the shell
   has solidified prior to break up (see \S 2.1 for the liquid case).
  A typical breaking strain
of iron is $\psi\sim 10^{-3}$ (e.g. Halliday \& Resnick \S 13.6).  The
 matter in the ejecta  expands  at the speed of sound ($c_{\rm s}$) 
  yielding a the break-up radius of,
\begin{equation}
r_{\rm b} = r_{\rm s}\left(1 + \psi 
\left[\frac{\beta c}{\Gamma c_{\rm s}}m_{\rm ejecta,-4}^{1/4} r^{-3/4}  \right]\right) 
\approx r_{\rm s}\left(1+\psi\right)\ ,
\end{equation}
where $r_{\rm s}$ is the solidification radius and $\beta c$ is the velocity of the ejecta.

The typical chunk size at birth is $c_{\rm s}\Delta t$
where $\Delta t = (r_{\rm b}-r_{\rm s})/c$, or, 
\begin{equation}
 \Delta r_{\rm c} \sim 186 \ {\rm cm}\  \psi_{-3} \rho_{b, 8}^{1/6}r_{\rm b,100} 
 =  186 \ {\rm cm}\  \zeta_{\rm  b} r_{\rm b,100}\ ,
\end{equation}
where we have defined $\zeta_{\rm b}=\psi_{-3}\rho_{\rm b,8}^{1/6}$.  
Here $\psi$ and the break-off radius are given in units
of $10^{-3}$ and 100 km respectively, while the density
at solidification radius is in units of $10^{8}$ g cm$^{-3}$.

The angle subtended (measured from the QS) by a broken chunk of the ejecta
at birth (i.e. at the break-up radius $r_{\rm b}$) is
\begin{equation}
   \theta_{\rm c} = \Delta r_{\rm c}/r_{\rm b}\sim 2 \times 10^{-5}\ {\rm rad}\ \zeta_{\rm b} \ .
\end{equation}
The corresponding solid angle is then,
\begin{equation}
  \Omega_{\rm c}=\pi \theta_{\rm c}^2 \sim 5\times 10^{-10}\ {\rm sr}\ \zeta_{\rm b}^{2}\ .
\end{equation}
The corresponding mass of the broken-off chunk is then roughly,
\begin{equation}\label{eq:mc}
\label{eq:chunkmass}
      m_{\rm c} =  \frac{\Omega_{\rm c}}{4 \pi} m_{\rm ejecta} 
      \sim 1.7\times 10^{19}\ {\rm gm}\  \zeta_{\rm b}^{2} m_{\rm ejecta, -4}\ .
\end{equation}
The ejecta break-up only depends on the stress in the two dimensions due to the ejecta's
lateral expansion being independent of its thickness. Thus, the chunk
mass is linearly dependent on the ejecta mass as a more massive ejecta
would be thicker. The total number of chunks can be estimated as,
\begin{equation}
 N_{\rm c}  =\frac{\theta_{\rm B, c}^2}{\theta_{\rm c}^2} 
 \sim 2.5\times 10^{7} \frac{\Gamma^{2} \theta_{\rm B, 0.1}^2}{\zeta_{\rm b}^{2}}\ ,
\end{equation}
where the collimation angle in the ejecta frame is $\theta_{\rm B,c}=\Gamma \theta_{\rm B}$.

\end{document}